\documentclass[preprint,showpacs,preprintnumbers,prb,amsmath,amssymb,aps,superscriptaddress]{revtex4-1}

\usepackage{graphicx}
\usepackage{dcolumn}
\usepackage{bm}
\usepackage{float}
\usepackage{subfig}
\usepackage{natbib}
\usepackage{xcolor}
\begin{document}

\title{Imaging uncompensated moments and exchange-biased emergent ferromagnetism in FeRh thin films}

\author{Isaiah Gray}
\affiliation{School of Applied and Engineering Physics, Cornell University, Ithaca, NY, 14853, USA}
\affiliation{Kavli Institute at Cornell, Cornell University, Ithaca, NY, 14853, USA}
\author{Gregory M. Stiehl}
\affiliation{Department of Physics, Cornell University, Ithaca, NY, 14853, USA}
\author{John T. Heron}
\affiliation{Department of Materials Science and Engineering, University of Michigan, Ann Arbor, Ann Arbor, Michigan, 48109, USA}
\author{Antonio B. Mei}
\affiliation{Department of Materials Science and Engineering, Cornell University, Ithaca, NY, 14853, USA}
\author{Darrell G. Schlom}
\affiliation{Department of Materials Science and Engineering, Cornell University, Ithaca, NY, 14853, USA}
\affiliation{Kavli Institute at Cornell, Cornell University, Ithaca, NY, 14853, USA}
\author{Ramamoorthy Ramesh}
\affiliation{Department of Materials Science and Engineering, University of California, Berkeley, CA, 94720, USA}
\author{Daniel C. Ralph}
\affiliation{Department of Physics, Cornell University, Ithaca, NY, 14853, USA}
\affiliation{Kavli Institute at Cornell, Cornell University, Ithaca, NY, 14853, USA}
\author{Gregory D. Fuchs}
\affiliation{School of Applied and Engineering Physics, Cornell University, Ithaca, NY, 14853, USA}
\affiliation{Kavli Institute at Cornell, Cornell University, Ithaca, NY, 14853, USA}

\date{\today}

\begin{abstract}

Uncompensated moments in antiferromagnets are responsible for exchange bias in antiferromagnet/ferromagnet heterostructures; however, they are difficult to directly detect because any signal they contribute is typically overwhelmed by the ferromagnetic layer. We use magneto-thermal microscopy to image uncompensated moments in thin films of FeRh, a room-temperature antiferromagnet that exhibits a 1st-order phase transition to a ferromagnetic state near 100~$^\circ$C. FeRh provides the unique opportunity to study both uncompensated moments in the antiferromagnetic phase and the interaction of uncompensated moments with emergent ferromagnetism within a relatively broad (10-15~$^\circ$C) temperature range near $T_C$. In the AF phase below $T_C$, we image both pinned UMs, which cause local vertical exchange bias, and unpinned UMs, which exhibit an enhanced coercive field that reflects exchange-coupling to the AF bulk. Near $T_C$, where AF and FM order coexist, we find that the emergent FM order is exchange-coupled to the bulk N\'eel order. This exchange coupling leads to the nucleation of unusual configurations in which different FM domains are pinned parallel, antiparallel, and perpendicular to the applied magnetic field before suddenly collapsing into a state uniformly parallel to the field. 
\end{abstract}

\maketitle

\section{Introduction}

In an ideal N\'eel antiferromagnet, each lattice-site spin is adjacent to an oppositely-pointing spin such that all spins are compensated \cite{NeelScience}. In real antiferromagnets, crystal defects \cite{MiltenyiPRL}, strain \cite{WhitePRL}, and surface roughness \cite{CharilaouJAP} (among other mechanisms) cause some spins to be uncompensated, resulting in local magnetic moments within the antiferromagnet \cite{ImryPRL, RohrerPRL}.
  
 Far from being a mere material imperfection, these uncompensated moments (UMs) are responsible for exchange bias in antiferromagnet (AF)/ferromagnet (FM) heterostructures \cite{NoguesJMMM, SchullerJMMM}, the most important current application of antiferromagnets due to its crucial role in pinning magnetoresistive sensors and other devices \cite{ReigSensors}. The exchange interaction between the AF and FM layers acts as an effective magnetic field, which shifts the $M(H)$ loop of the FM along the horizontal field axis \cite{MeiklejohnPhysRev}. The underlying mechanisms of exchange bias are complex: The interfacial FM spins are not coupled to the entire AF surface as initially thought \cite{JungblutJAP} but instead to UMs which comprise a small percentage of the AF surface \cite{TakanoPRL, BruckPRL}. Surprisingly, experimentally altering the bulk AF domain structure -- for example by introducing bulk defects via irradiation \cite{MiltenyiPRL} - affects the magnitude of EB, even if the AF/FM interface is unchanged. Therefore, bulk UMs must also contribute to exchange bias.
 
 After decades of intensive study, the domain state model \cite{NowakPRB, KellerPRB} has emerged as the generally accepted model of exchange bias. In this model, bulk AF domains acquire uncompensated moments upon field-cooling from above the N\'eel temperature $T_N$, both at AF domain walls and within the domains themselves. These bulk AF UMs stabilize the interfacial UMs that directly exchange-couple to the FM spins to cause exchange bias. Despite their critical role in exchange bias, bulk UMs are difficult to study directly: any signal they produce is typically overwhelmed by the FM layer in exchange-biased bilayers, while the exchange pinning between the UMs and the AF bulk makes the UMs in a single AF layer difficult to manipulate. Therefore, a detailed experimental understanding of how the spatial structure of the bulk UMs stabilizes the interfacial UMs is still lacking \cite{SchullerJMMM}.
 
 The metallic AF FeRh offers a potential path to circumvent the difficulties of directly studying UMs. An antiferromagnet at room temperature, it undergoes an unusual 1st-order phase transition from AF to FM near $100~^\circ$C \cite{McKinnonJPhysC}. This transition is interesting in itself -- the exact mechanism is still debated \cite{WollochPRB, LewisJPhysD}  -- and is also exploited in potential electric field-assisted \cite{LiuPRL, ZhengSciRep, CherifiNatMater} and heat-assisted \cite{ThieleIEEETransac} magnetic recording devices. Within the relatively broad (10-15~$^\circ$C) transition region, AF and FM phases coexist and interact. The phase transition allows detection of UMs in the AF phase, where they are not overwhelmed by FM moments, and then study of the interaction between UMs and emergent ferromagnetism within the transition region. 
  
 In this work we image uncompensated moments and emergent FM in FeRh using magneto-thermal microscopy \cite{BartellNatComm, BartellPRAppl, GuoPRB, GuoPRAppl, WeilerPRL, GrayArxiv, IguchiArxiv}. While most magnetic imaging techniques are primarily surface-sensitive \cite{KappenbergerPRL, BruckPRL}, magneto-thermal microscopy is based on through-plane thermal gradients within the AF thin film and therefore can potentially resolve bulk UMs. Below the transition temperature $T_C$, we resolve pinned and unpinned UMs within the AF film, which result in \textit{vertical} exchange bias that locally shifts the entire $M(H)$ loop above or below zero. Near $T_C$, we find that the emergent FM phase is exchange-coupled to the bulk AF order. Our images reveal a disordered exchange-biased AF/FM system in which different FM domains are pinned parallel, antiparallel, and perpendicular to the applied magnetic field, even when the applied field is much greater than the coercivity in the FM phase. These results demonstrate previously unobserved exchange bias within a metamagnetic phase transition and suggest a general method for spatially resolving uncompensated moments in AF metals with magneto-thermal microscopy.

\section{Imaging uncompensated moments in F\lowercase{e}R\lowercase{h} in the AF phase}

\subsection{Experimental setup and materials}

Magneto-thermal microscopy is based on the anomalous Nernst effect (ANE) \cite{BartellNatComm, GuoPRB, GuoPRAppl, WeilerPRL, IguchiArxiv} and the longitudinal spin Seebeck effect (LSSE) \cite{BartellPRAppl, GrayArxiv}. In the anomalous Nernst effect, a thermal gradient $\bm{\nabla} T$ in a magnetic conductor with moment $\bm{m}$ produces an electric field $\bm{E} = -N \mu_0 \bm{\nabla} T \times \bm{m}$. In the longitudinal spin Seebeck effect, a thermal gradient $\bm{\nabla} T$ within a magnetic material generates a pure spin current $\bm{j}_s \parallel \bm{\nabla}T$. If the magnetic material is interfaced with a heavy metal, some of $\bm{j}_s$ diffuses into the heavy metal where it is transduced by the inverse spin Hall effect into a charge current $\bm{j}_c \perp \bm{j}_s$. ANE and LSSE have the same symmetry: in both cases, $\bm{\nabla}T$ in a patterned device produces an overall voltage drop proportional to the in-plane magnetic moment. 

We image epitaxial MgO(001)/20 nm $\mathrm{Fe}\mathrm{Rh}$(001)/10 nm Pt, patterned into 3~$\mu$m$\times$18~$\mu$m Hall crosses by photolithography and ion milling. The FeRh is sputtered from a stoichiometric $\mathrm{Fe}_{0.49}\mathrm{Rh}_{0.51}$ target; from x-ray diffraction data in Appendix A and Vegard's law, we estimate the composition of the film to be $\mathrm{Fe}_{0.47}\mathrm{Rh}_{0.53}$. Because FeRh is metallic in both the AF and FM phases and because Pt is a heavy metal, both ANE and LSSE \cite{WuPRL, SekiPRL} can contribute to our signal. Additional magneto-thermal images of uncapped 35 nm-thick MgO(001)/$\mathrm{Fe}_{0.52}\mathrm{Rh}_{0.48}$ and MgO(001)/$\mathrm{Fe}_{0.43}\mathrm{Rh}_{0.57}$ shown in Appendix D yield similar signal magnitudes as the 20 nm-thick $\mathrm{Fe}_{0.49}\mathrm{Rh}_{0.51}$/Pt, which indicates that a potential LSSE signal is smaller than the signal contribution from ANE \footnote{In addition to FM LSSE from uncompensated moments, both bulk and interfacial AF LSSE are possible. However, we expect signal from AF LSSE to be linear in field, which is not present in the data. We can rule out interfacial AF LSSE because it relies on an uncompensated AF interface and the interface of FeRh(001) is compensated}. Therefore the Pt layer does not affect our conclusions, and for convenience we refer to ANE plus a potential smaller FM LSSE as ANE.

 We generate local thermal gradients using 3-ps-pulses from a 780 nm-wavelength Ti:Sapphire laser focused to a 650 nm-diameter spot. We raster scan the laser over the device and detect the resulting $V_{ANE}$ pulses using a time-domain homodyne technique described in detail previously \cite{BartellNatComm}. Note that $V_{ANE}$ represents a weighted average of the in-plane $\bm{m}$ within the local out-of-plane thermal gradient, $\nabla T_z$. Further details are available in Appendix G.   

\begin{figure}[htb]
\centering
\includegraphics[scale=0.81]{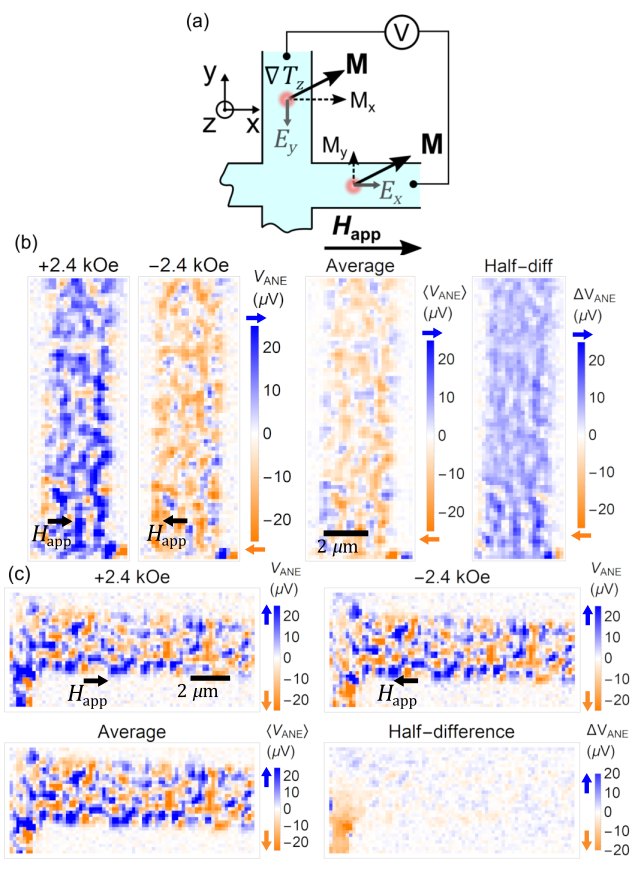}
\caption{Anomalous Nernst imaging of MgO(001)/20 nm $\mathrm{Fe}_{0.49}\mathrm{Rh}_{0.51}$/6 nm Pt Hall crosses at 25 $^\circ$C. (a) Schematic of the measurement. We make electrical contact to the crosses in an L-shape, therefore we measure $m_x$ in the vertical branch and $m_y$ in the horizontal branch. (b) The vertical branch at $H_{app}$ = $\pm$2.4 kOe along $x$. The average between images at positive and negative field shows pinned uncompensated moments that are unaffected by field, while the half-difference shows unpinned uncompensated moments that reverse with field. (c) Imaging the horizontal branch. Only pinned moments appear in the ANE image, since the unpinned moments rotate along $x$ to be parallel to $H_{app}$.}
\end{figure} 

\subsection{Pinned and unpinned uncompensated moments}

We first image FeRh/Pt at room temperature, in the AF phase, as a function of applied magnetic field $H_{app}$. Fig.~1 shows ANE images of a 3 $\mu$m cross at $H_{app}$~=~$\pm$2.4 kOe along the $x$-direction. To probe both $x$ and $y$-components of the magnetization, we make contact to the Hall cross in an L-shape, illustrated in Fig. 1(a). $V_{ANE}$ is proportional to the in-plane component of $\bm{m}$ locally perpendicular to the device channel. Therefore, we measure $m_x$ (collinear with the magnetic field) in the vertical branch and $m_y$ (perpendicular to the magnetic field) in the horizontal branch. In the vertical branch in Fig.~1(b) we observe micron-scale regions of positive and negative contrast which partially switch with field. Unlike anomalous Nernst images of ferromagnets \cite{BartellNatComm, GuoPRAppl, GuoPRB}, $V_{ANE}$ does not uniformly saturate with the field, which indicates that it does not originate from simple ferromagnetism. We can rule out possible spurious contributions from spatial inhomogeneity in sample resistivity or thermal conductivity, for two reasons. First, spatially inhomogeneities produce characteristic dipole-like patterns in the $V_{ANE}$ images from the charge Seebeck effect, which we do not observe \cite{MeiPrep}. Second, the inhomogeneous contrast disappears above the transition temperature $T_C$ (see Appendix B). The images are reproducible in detail upon repeated heating and cooling cycles (see Appendix E).

 \begin{figure}[htb]
\includegraphics[scale=0.36]{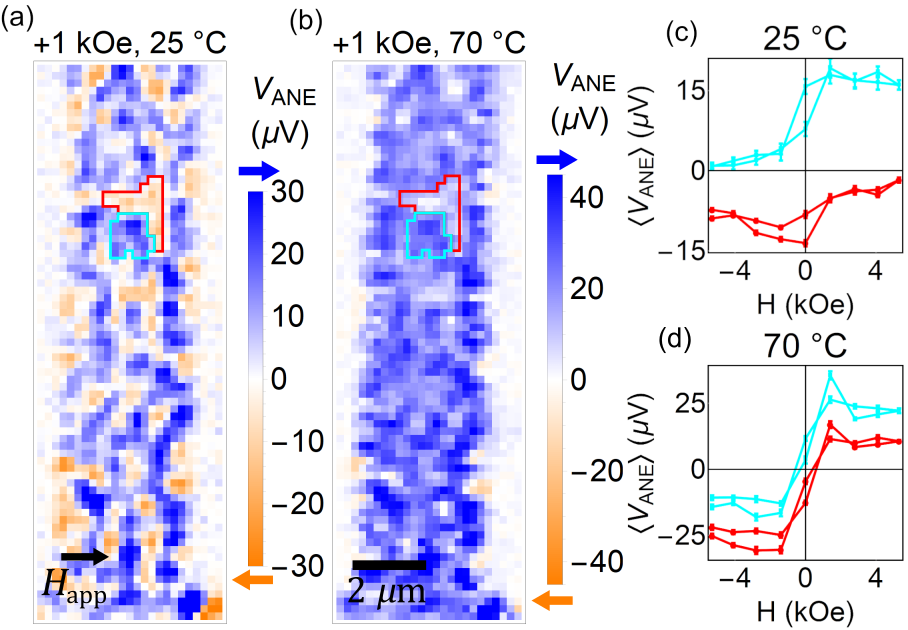}
\caption{Temperature dependence of the uncompensated moments in the AF phase. (a, b): ANE images of the 3-$\mu$m-wide Hall cross in Fig. 1, at (a) 25~$^\circ$C and (b) 70~$^\circ$C, at $H_{app}$ = +1 kOe $\parallel~x$. Green and red line outline adjacent AF domains, each containing a net \textit{pinned} moment. (c,d) Average $V_{ANE}$ of all pixels within the cyan and red outlined domains as a function of $H_{app}$, at 25~$^\circ$C (c) and 70~$^\circ$C (d). Ferromagnetic hysteresis loops with vertical shifts show \textit{vertical} exchange bias from exchange coupling between the UMs and the bulk N\'eel order. At 70~$^\circ$C the coercivity and the magnitude of the vertical shifts decrease as more pinned UMs become unpinned. }
\end{figure} 

 We distinguish the portion of the signal that switches with $H_{app}$~=~$\pm~2.4$ kOe from the portion that does not by taking the half-difference and average, respectively. The half-difference switches at $H_{app}$ = 1 kOe, which is consistent with VSM measurements in Appendix C of a weak residual moment with about 750 Oe coercive field. Meanwhile, the average $V_{ANE}$ between positive and negative field is unaffected up to $H_{app}$~=~$\pm$5.2 kOe, the largest field we can apply in our setup.  Fig.~1(c) shows that the signal in the horizontal branch does not switch, as expected, because the moments that switch align with $H \parallel\hat{x}$, whereas we measure $m_y$ here.

We attribute the micron-size regions of positive and negative $V_{ANE}$ in the average ANE image in Fig.~1(b) and the ANE images in Fig.~1(c) that do not switch with field to pinned uncompensated moments that are strongly coupled to the bulk N\'eel order. By definition pinned UMs carry a magnetic moment, therefore they should contribute an anomalous Nernst signal. In addition, the contrast disappears above the transition temperature $T_C$, which rules out possible spurious contributions from spatial inhomogeneity in sample resistance or thermal conductivity.

 According to the domain state model \cite{NowakPRB}, pinned UMs occur both in the bulk and at the interfaces of the AF. They can arise either within AF domain walls, or within AF domains from an Imry-Ma-type statistical imbalance in the number of defects in each of the two spin sublattices \cite{ImryPRL}. We can rule out AF domain walls as the dominant source of $V_{ANE}$, because AF domain walls are typically tens of nanometers wide \cite{BaldasseroniJPhysCondMat} and would not be resolvable with our 650-nm resolution.  In the Imry-Ma mechanism each AF domain carries a small net magnetization collinear with the N\'eel order. AF domains in FeRh thin films range between 300 nm and 2 $\mu$m in size depending on the defect density and growth methods \cite{BaldasseroniJPhysCondMat}, which is consistent with the 1-2~$\mu$m domains in the $V_{ANE}$ images in Fig.~1(b) and 1(c). Nevertheless, we cannot directly confirm that $V_{ANE}$ originates from magnetized AF domains without corresponding XMLD-PEEM images of the AF domains of the same sample, which suggests a direction for further experiments.

The contrast in the half-difference image in Fig.~1(b), representing moments that switch with magnetic field, could originate from unpinned UMs in the AF bulk and interfaces \cite{KandeJAP}, or from an interfacial residual FM phase distinct from the AF bulk, which is common in FeRh thin films. Residual FM can occur at both the top and bottom interface \cite{GatelNatComm} and has been variously attributed to strain \cite{FanPRB}, surface symmetry breaking \cite{PressaccoSciRep}, and chemical diffusion \cite{BaldasseroniJAP}. All these sources of residual FM signal listed could contribute to $V_{ANE}$. However, previous studies of the effects of capping layers found no residual FM phase at the compensated FeRh(001)/Pt interface \cite{BaldasseroniJAP}, which suggests that if there is a residual FM phase it most likely occurs at the bottom rather than the top interface. Regardless, its presence does not significantly affect our conclusions. 
 
  To investigate the temperature dependence of pinned and unpinned UMs below $T_C$, we image the sample in Fig.~1 as a function of applied field $H_{app}$ at 25~$^\circ$C and 70~$^\circ$C. Example images at $H_{app} = +1$ kOe are shown in Fig.~2(a) and 2(b). We identify two adjacent magnetized AF domains, outlined in red and cyan line, using the zero crossing at zero magnetic field as the perimeter. We compute the average $V_{ANE}$ of all pixels within these two domains for each value of $H_{app}$. We plot these averages as a function of $H_{app}$ at 25~$^\circ$C in Fig.~2(c) and at 70~$^\circ$C in Fig.~2(d). At 25~$^\circ$C, we obtain ferromagnetic hysteresis loops that are vertically shifted, enough to move the entire loop above or below zero, while at 70~$^\circ$C the contrast is more uniform and the magnitude of the vertical shifts decreases from 10~$\mu$V to $\sim 4 ~\mu$V.
  
\begin{figure*}
\includegraphics[scale=0.44]{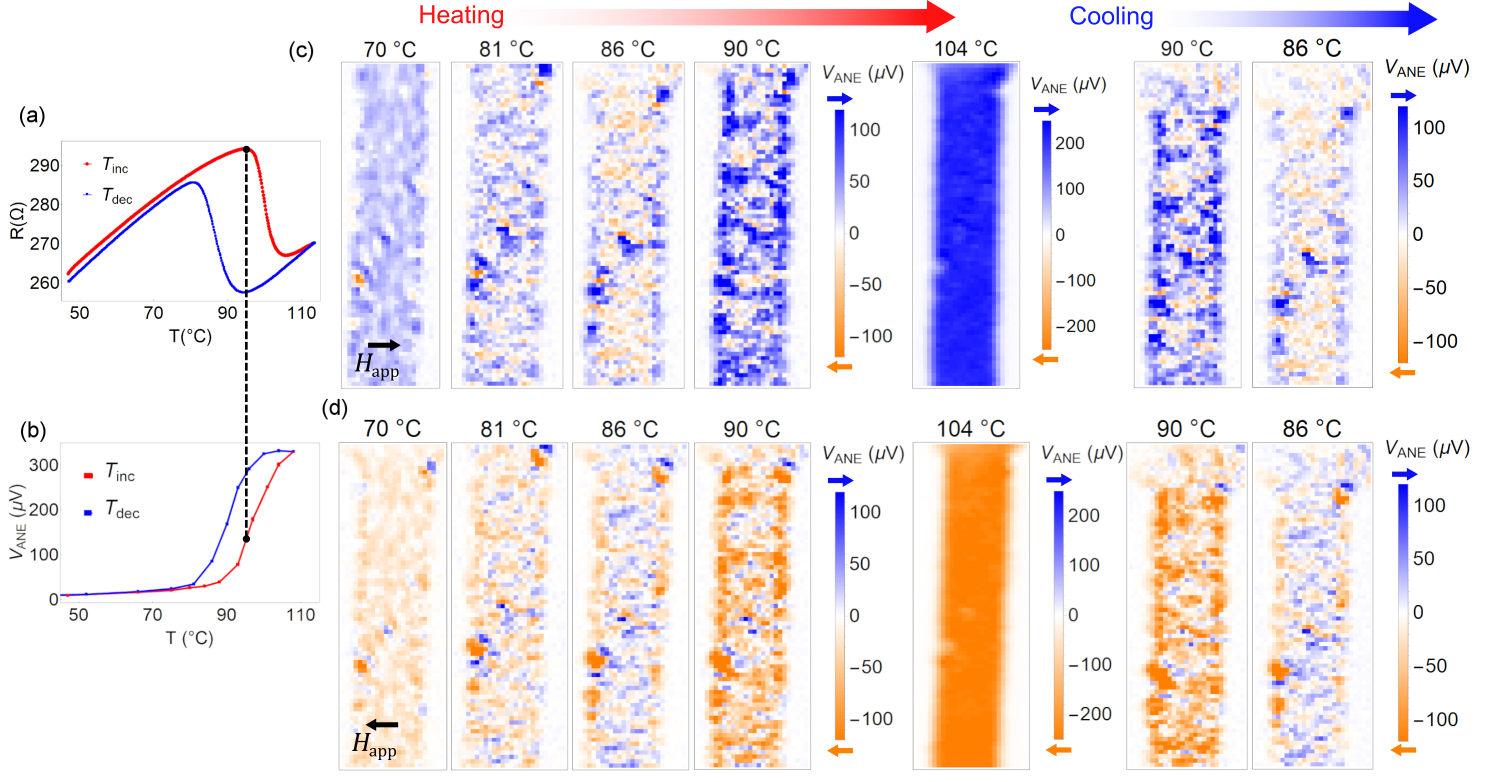}
\caption{ Imaging the metamagnetic phase transition from AF to FM. (a) Characterization of the critical temperature $T_C$ using decreased electrical resistivity in the FM phase. (b) $V_{ANE}$ at a single position as a function of temperature, showing similar hysteresis. (c) ANE imaging of the transition at $H_{app}$ = +2 kOe applied along $+x$. We observe nucleation, percolation, and growth of FM domains, characteristic of the 1st-order transition. Surprisingly, the emergent FM domains at 86~$^\circ$C are oriented antiparallel to $H_{app}$, indicating that they are exchange-coupled to the bulk N\'eel order. (d) Imaging the transition again at $H_{app}$ = -2 kOe. Again the FM domains nucleate oriented antiparallel to $H_{app}$, which suggests that the exchange-coupling $J_{F-AF}$ between emergent FM and bulk AF is antiferromagnetic.}
\end{figure*} 
  
  Temperature-dependent vertical shifts and coercivity enhancement are the experimental signatures of \textit{vertical} exchange bias \cite{ZhengJAP, PassamaniJMMM}, which is less common than the usual horizontal exchange bias in AF/FM bilayers. In horizontal EB, the moment originates from the FM and the horizontal shift yields the effective field from exchange coupling to the AF. In vertical EB, however, the vertical shift directly reflects the moment from the pinned UMs. Vertical EB is rarely observed experimentally, first because the net moment from pinned UMs in the AF layer is typically much smaller than the moment from the FM layer and second because the pinned UMs are not uniformly oriented (as Fig.~1 shows) and the net moment averages to nearly zero over many domains. In our images there is no large FM background, therefore we are able to resolve relatively weak moments from pinned UMs ($V_{ANE}$ at 25~$^\circ$C is about 15 times smaller than in the full FM phase at 104~$^\circ$C). The enhanced coercivity of the unpinned UMs -- 750 Oe at 25~$^\circ$C compared to 50 Oe in the FM phase -- shows that the unpinned UMs are exchange-coupled to the AF bulk. This is consistent with a modified Stoner-Wolfarth model of uncompensated moments \cite{HennePRB, BuchnerPRB} in which varying degrees of exchange-coupling result in fully rotatable (unpinned) UMs, partially pinned UMs which enhance the coercivity, and fully pinned UMs which do not rotate at all with magnetic field. At 70~$^\circ$C some of the pinned moments become unpinned, resulting in decreased vertical shifts.  

\section{Interaction of coexisting antiferromagnetic and ferromagnetic phases near $T_C$}

 We next perform ANE imaging as a function of temperature and magnetic field through the phase transition, in which AF and FM order can coexist. We first characterize the transition in Fig.~3(a) through the drop in resistivity from magnetoresistance \cite{KouvelJAP}, which shows the temperature hysteresis characteristic of a 1st-order phase transition. In Fig. 3(b), we plot the ANE voltage from one point on the sample at $H_{app}$ = 2.0 kOe on the same temperature scale, which shows similar hysteresis in the magnetic moment. The local $T_C$ probed by the laser spot in Fig.~3(b) is about 8~$^\circ$C lower than $T_C$ measured by the resistivity in Fig.~3(a), which may be due both to spatially varying $T_C$ and a lower effective $T_C$ from laser heating \footnote{At the 3~$\mathrm{mJ}/\mathrm{cm}^2$ fluence employed, we estimate a peak temperature increase of $\sim$25~$^\circ$C (see Appendix G). The fact that $T_C$ measured with ANE is 8~$^\circ$C lower than the bulk $T_C$ measured with resistivity may be due to a higher $T_C$ required to nucleate a submicron FM domain in an AF background than to uniformly heat the entire sample}. We then image through both heating and cooling portions of the transition, first at $H_{app}$ = +2 kOe and then $H_{app}$ = -2 kOe applied along the $x$-axis, which are shown in Fig.~3(c) and Fig.~3(d), respectively. Note the data in Fig. 3(a) and 3(b) are taken on different devices as the images in Fig. 3(c) and Fig. 3(d), although they are fabricated from the same film. 

The ANE images through the transition in Fig.~3(c) show that the FM phase nucleates, percolates, and coalesces, in agreement with previous imaging studies in FeRh\cite{BaldasseroniAPL, BaldasseroniJPhysCondMat, GatelNatComm, AlmeidaSciRep}. At 70~$^\circ$C the contrast shows UMs and residual FMs in the AF phase. We observe FM domains nucleating first at sample edges and defects at 81~$^\circ$C and 86~$^\circ$C, percolating through the device at 90~$^\circ$C until $V_{ANE}$ is nearly uniform in the FM phase at 104~$^\circ$C. Unexpectedly, we find that some of the FM domains nucleate with an orientation that is not parallel to the applied field, even though $H_{app}$ = 2~kOe is much greater than the 50~Oe coercivity field in the FM phase. At 86 $^\circ$C in Fig.~3(c), the moments on the edges tend to have positive $m_x$ (blue), parallel to $H_{app}$, while the moments near the center of the device tend to have \textit{negative} $m_x$ (orange), \textit{antiparallel} to $H_{app}$. At $H_{app}$~=~-2~kOe in Fig.~3(d) we observe the same phenomenon: the orange moments at the edges tend to be parallel to $H_{app}$ and the blue moments in the middle tend to be antiparallel to $H_{app}$. As the FM moments coalesce above 90~$^\circ$C, they reorient to be parallel to the applied field for both orientations of $H_{app}$. The spatial structure in the heating branch reproduces in the cooling branch, seen by comparing the corresponding heating and cooling images at 90 $^\circ$C and 86 $^\circ$C.

\begin{figure}[htb]
\centering
\includegraphics[scale=0.45]{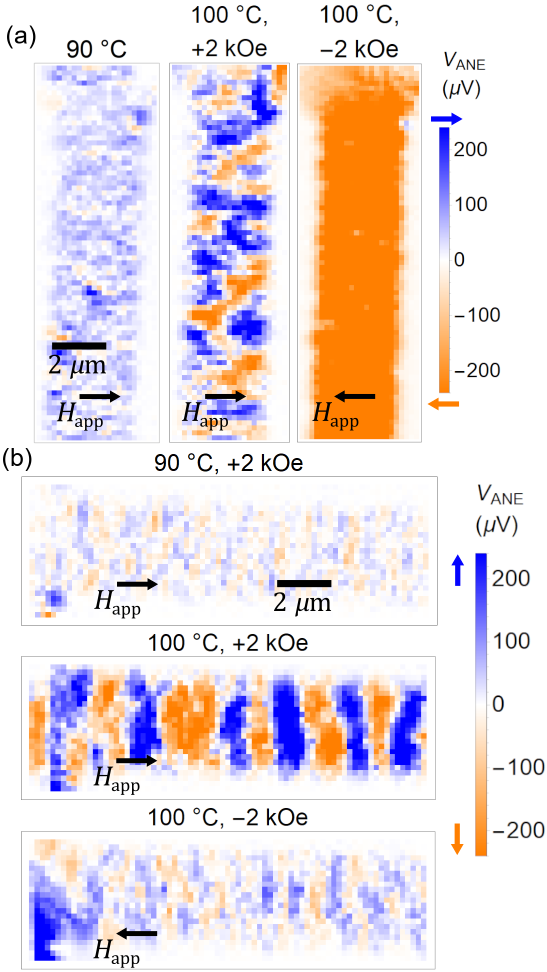}\caption{Metastable states of of FM domains pinned parallel, antiparallel, and perpendicular to $H_{app}$ = 2~kOe applied field. (a) In the same Hall cross as in Fig.~3, we observe emergent FM domains pinned parallel and antiparallel to the AF order at 100~$^\circ$C. which collapse into the uniform FM state after reversing the direction of $H_{app}$. (b) Corresponding images of the horizontal branch of the same cross, acquired simultaneously as (a). At 100~$^\circ$C and $H_{app}$ = +2~kOe we observe large-scale emergent FM domains oriented perpendicular to $H_{app}$. Weak contrast at $H_{app}$~=~-2~kOe shows that the magnetic structure collapses into the uniform FM state, oriented along $-x$.}
\end{figure}  

 We explain these puzzling results in terms of exchange bias between FM interfaces and the AF bulk within the FeRh film near $T_C$. Previous cross-sectional imaging of the phase transition in FeRh with electron holography showed that the transition occurs first at the top and bottom interfaces before spreading into the bulk \cite{GatelNatComm}. Therefore, for some temperature range near $T_C$ the interfaces of the FeRh film are FM while the bulk is still AF, forming a kind of AF/FM heterostructure with rough interfaces. Our observation of newly nucleated FM domains that are not parallel to $H_{app}$, even when $H_{app}$~=~2~kOe is greater than both the 50~Oe coercivity field in the FM phase and the $\sim$750~Oe coercivity field of the unpinned UMs in the AF phase, indicates that the FM interfaces are pinned by exchange-coupling to the pinned UMs; in other words, the FM interfaces are exchange-biased by the AF bulk as previously suggested by spin-wave resonance measurements in Pd-doped FeRh \cite{MasseyArxiv}.
 
 \begin{figure}[htb]
\centering
\includegraphics[scale=0.47]{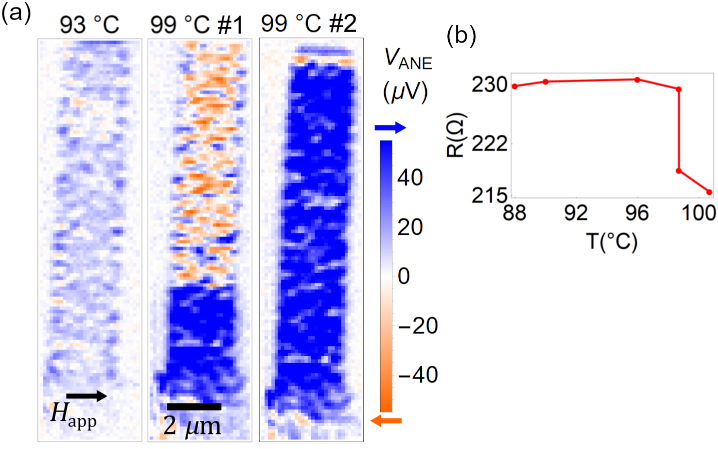}
\caption{Spatially inhomogeneous collapse of metastable exchange-coupled states on a similar device as in Fig.~4. At 93~$^\circ$C and $H_{app}$ = 2.0 kOe along $+x$ we image unpinned UMs and residual FM. At 99~$^\circ$C the lower portion of the cross is uniformly FM parallel to $H_{app}$, while the upper portion is an exchange-biased mixture of AF and FM with antiparallel exchange coupling. Immediately imaging again causes the FM to collapse uniformly parallel to $H_{app}$, suggesting an avalanche-like transition proceeding by front propagation. (d) Sample resistance during the imaging in (c). The resistance decreases suddenly at the point of the collapse in magnetic structure.}
\end{figure} 
 
To observe both $m_x$ and $m_y$ near $T_C$, in Fig.~4(a) and 4(b) we image the vertical and horizontal branch of the same device as in Fig.~3. At 90~$^\circ$C and $H_{app}$ = +2 kOe magnetic field applied along the $x$-direction, we observe mostly positive $V_{ANE}$ in the vertical branch and weak contrast in the horizontal branch. $V_{ANE}$ at this temperature represents unpinned UMs parallel to $H_{app}$. At 100~$^\circ$C and $H_{app}$ = 2~kOe, however, we observe the sudden emergence of large-scale ($>2~\mu$m) FM domains of positive and negative $V_{ANE}$ in both the vertical and horizontal branch. Because we measure $m_x$ in the vertical branch and $m_y$ in the horizontal branch, this data shows that some FM domains are parallel, others antiparallel, and still others perpendicular to $H_{app}$, all coexisting simultaneously. We interpret the different orientations of the FM domains with respect to $H_{app}$ to be manifestations of different exchange biases. Depending on the mechanism, the exchange coupling between the FM moments and the pinned UMs can be parallel\cite{NoguesPRL}, antiparallel \cite{GrimsditchPRL}, or perpendicular \cite{MaatPRL}, and different directions of exchange bias can coexist simultaneously \cite{OhldagPRL2006, RaduPRB}.

 Upon reversing the direction of $H_{app}$, the FM domains abruptly collapse to be parallel to $H_{app}$, which appears as nearly uniform $V_{ANE}$ in the vertical branch and weak contrast in the horizontal branch. This sudden collapse may indicate that the states of antiparallel and perpendicular exchange coupling are unstable to perturbations in field. In addition, because applying a magnetic field reduces $T_C$ with a slope of 9~K/T \cite{MaatPRB}, $T_C$ is effectively raised and then lowered by 1.8~$^\circ$C upon reversing the 2~kOe field, therefore the sudden collapse may also reflect instability of the exchange coupling to temperature.
 
 Imaging another 3~$\mu$m-wide device from the same film at $H_{app}$ = 2.0 kOe applied field in Fig.~5(a) while measuring the resistance in 5(b) shows even more puzzling behavior. At 93~$^\circ$C we observe unpinned UMs in the AF phase, oriented parallel to $H_{app}$. At 99~$^\circ$C, maintaining $H_{app}$ at +2.0 kOe, we see that the moments in the upper portion of the channel are mostly pinned antiparallel to $H_{app}$, while the moments in the lower third are parallel to $H_{app}$. Upon immediately retaking the image, the entire sample has collapsed into the FM phase parallel to $H_{app}$, accompanied by a sharp decrease of the resistance in Fig.~5(b) which indicates the transition into the FM phase. This data is reproducible as shown in Appendix F, and additionally because the raster scanning proceeds from bottom to top, the spatial phase inhomogeneity is not due to the sample collapsing into the FM phase during the 20-30 min process of imaging. The spatial inhomogeneity and sudden collapse of the exchange-biased AF/FM heterostructure shows that the states of antiparallel and perpendicular exchange coupling are metastable and suggests (but does not prove) that in this case the transition is avalanche-like \cite{CarrilloPRB} and proceeds by front propagation \cite{ShawActaMaterialia}.  

Although we do not have a comprehensive understanding of the mechanisms behind this complicated metastable pattern of simultaneous parallel, antiparallel, and perpendicular exchange coupling in FeRh, we can gain some insight by comparing to antiparallel and perpendicular exchange coupling in conventional AF/FM bilayers.   In antiparallel EB (usually called \textit{positive} EB) \cite{NoguesPRL}, the exchange coupling $J_{F-AF}$ between interfacial FM spins and pinned AF UMs is antiferromagnetic \cite{NoguesPRB, GredigAPL}. The pinned UMs are unpinned by heating above the blocking temperature $T_B$ and set parallel to the cooling field $H_{FC}$, either by applying a magnetic field large enough to overcome $J_{F-AF}$ \cite{FitzsimmonsPRB2007} or by training through repeated field reversals \cite{MishraPRL, RaduPRB}. After removing $H_{FC}$, $J_{F-AF}$ rotates the FM spins antiparallel to the pinned UMs and hence to $H_{FC}$. Perpendicular exchange coupling, where the FM easy axis is orthogonal to the N\'eel orientation, is associated with either a slight canting of the AF sublattices \cite{MoranAPL} or a rough AF/FM interface in which the collinear exchange interaction $J_{F-AF}$ is frustrated by the energy required to create FM domain walls \cite{MalozemoffPRB}.

To determine if the mechanisms underlying the FM spins pinned antiparallel to $H_{app}$ in our ANE images are similar to the ones that contribute to positive exchange bias in AF/FM bilayers, we repeatedly image the phase transition while varying $H_{FC}$, the applied field $H_{app}$, and the laser fluence $F$ (shown in Appendix F). Varying $H_{FC}$ from +2 kOe to -2 kOe before imaging does not significantly alter the antiparallel exchange-coupling, nor does changing $H_{app}$ from 2 kOe to 100 Oe. This means that the orientation of the pinned AF moments in our ANE images is not set by field-cooling prior to imaging, nor is it set by applying a large $H_{app}$ to overcome $J_{F-AF}$ during imaging. Surprisingly, we do not observe FM moments antiparallel to $H_{app}$ when using $F$ = 0.6 $\mathrm{mJ}/\mathrm{cm}^2$ instead of 2.0 $\mathrm{mJ}/\mathrm{cm}^2$, which indicates that heating from the pulsed laser has a different effect than adiabatic heating of the whole sample with the background heater. We speculate that laser heating initially unpins the pinned AF UMs, and after repeated heating and cooling they become pinned parallel to $H_{FC}$ and rotate the emergent FM spins antiparallel to $H_{FC}$. The process may be similar to the training-induced positive exchange bias mentioned earlier \cite{MishraPRL}, which appears only after several training cycles near $T_N$. Further studies varying the number of pulses delivered to each pixel may be necessary to confirm this hypothesis.

\section{Conclusion}

In summary, we use anomalous Nernst microscopy to image uncompensated moments in the AF phase of FeRh below $T_N$ as well as emergent FM near $T_N$. We resolve enhanced coercivity and spatially inhomogeneous vertical exchange bias below $T_N$, demonstrating varying degrees of exchange-coupling between UMs and the bulk N\'eel order.  In addition, we demonstrate that newly nucleated FM domains near $T_N$ are exchange-coupled to the pinned UMs even in the presence of a nominally saturating magnetic field, providing a direct experimental demonstration of exchange bias within a single FeRh thin film. We expect the imaging of uncompensated moments with anomalous Nernst microscopy to extend to a variety of AF metals, which could lead to a better understanding of the role of bulk UMs on exchange bias in both pure AFs and AF/FM bilayers.

\section{Acknowledgments}
 
This research was primarily supported by the Cornell Center for Materials Research with funding from the NSF MRSEC program (DMR-1719875). This work made use of the CCMR Shared Facilities and the Cornell NanoScale Facility, an NNCI member supported by NSF Grant ECCS-1542081. A.B.M. and D.G.S. acknowledge support in part by the Semiconductor Research Corporation (SRC) under nCORE task 2758.003 and by the NSF under the E2CDA program (ECCS-1740136). G.M.S. acknowledges support by the National Science Foundation (DMR-1708499). Materials synthesis was performed in part in a facility supported by the National Science Foundation [Platform for the Accelerated Realization, Analysis, and Discovery of Interface Materials (PARADIM) under Cooperative Agreement No. DMR-1539918.

\bibliographystyle{apsrev4-1}
\bibliography{FeRh_bib}

\appendix

\section{FeRh-Pt growth and characterization}

The FeRh/Pt thin films in the main text were grown by DC sputtering from a stoichiometric $\mathrm{Fe}_{0.49}\mathrm{Rh}_{0.51}$ target onto single-crystal MgO(001) substrates. The base pressure was $2\times10^{-8}$ torr. The samples were grown at 375~$^\circ$C and annealed at 520~$^\circ$C for 1 hour.

\begin{figure*}[htb]
\centering
\includegraphics[scale=0.38]{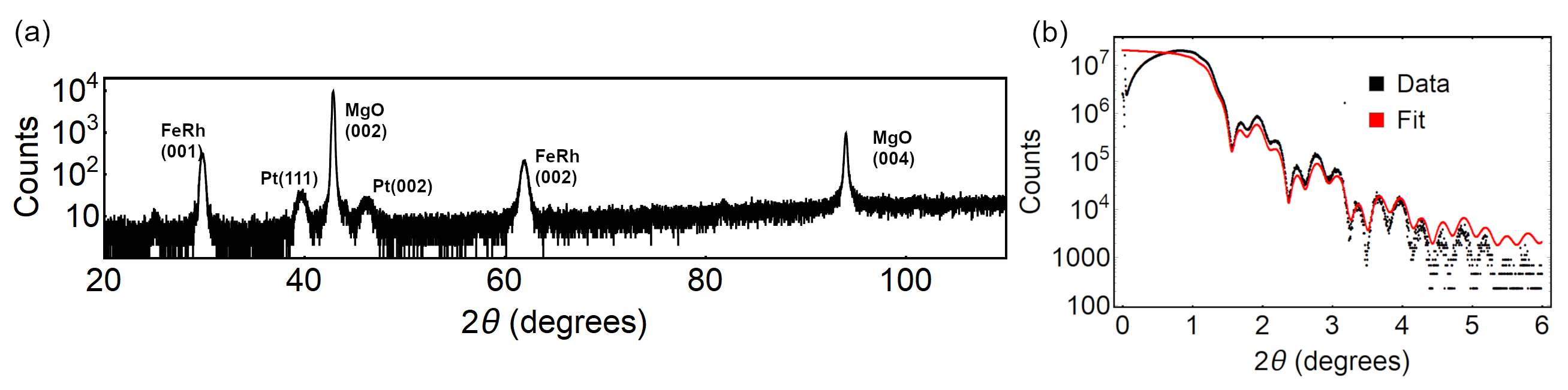}
\caption{Structural characterization of the MgO(001)/FeRh/Pt samples. (a) XRD scan demonstrating epitaxial growth of FeRh(001) on MgO(001). (b) XRR scan of the same film. From the fit we estimate the thicknesses of the FeRh and Pt layers to be 20.5 nm and 8.0 nm, respectively.}
\label{growth}
\end{figure*}

Fig.~\ref{growth}(a) shows the XRD scan of the MgO(001)/FeRh/Pt samples imaged in the main text. The FeRh(001) and (002) peaks demonstrate epitaxial growth of B2 CsCl FeRh(001). Previous structural characterization of ordered B2 CsCl FeRh as a function of growth composition \cite{MeiAPL} showed a linear relation between the Rh concentration and the strain in the AF phase, manifesting in the FeRh peak positions in the XRD. Using this linear relation we estimate the Fe and Rh concentrations to be 47$\%$ and 53$\%$, respectively. 

From the FeRh(002) peak at $2\theta = 61.9^\circ$ we obtain an out-of-plane lattice constant of $a = 0.2996$ nm. This value is greater than the bulk value of $0.2986$ nm for $\mathrm{Fe}_{1-x}\mathrm{Rh}_x$ with $52~<x~<60$ \cite{ShiranePhysRev}, which indicates that the FeRh is compressively strained. Note that because the AF-FM phase transition is accompanied by a $\sim1\%$ lattice expansion, the strain is greater in the FM phase than in the AF phase. This rules out the possibility that the contrast we attribute to pinned UMs in Fig.~1 and Fig.~2 of the main text is due to strain, and that the contrast disappears in the FM phase because the lattice expansion relaxes the strain. 

 Fig.~\ref{growth}(b) shows XRR data on the same FeRh/Pt film. From the fit we estimate the thicknesses of the FeRh and Pt layers as 20.5 nm and 8.0 nm,  respectively, and the surface roughness to be 0.46 nm for both layers.

\section{Disappearance of AF uncompensated moments in the FM phase}

\begin{figure*}[!htb]
\centering
\includegraphics[scale=0.55]{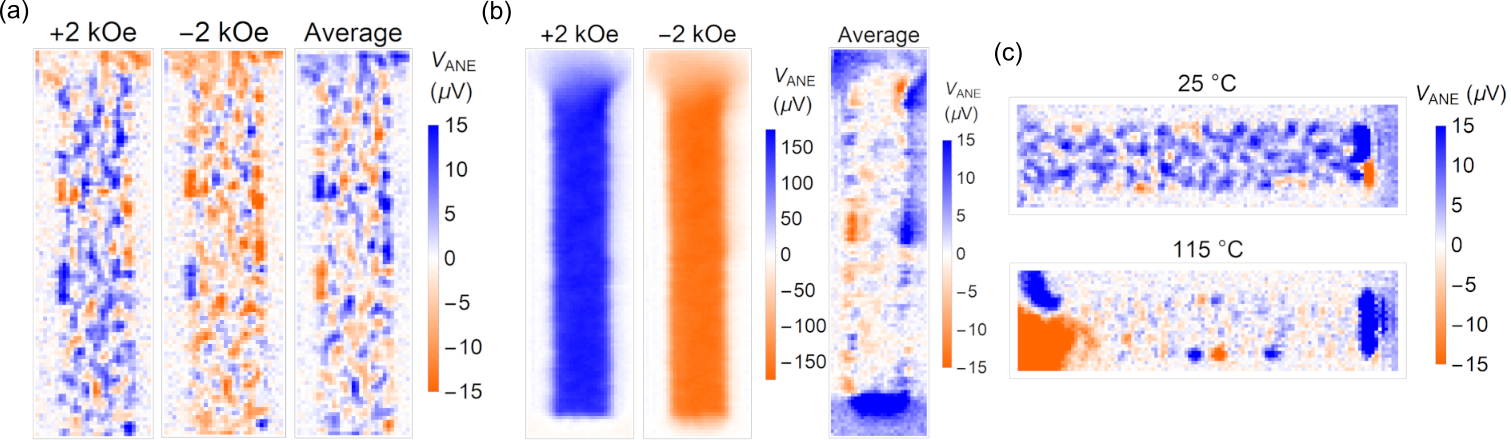}
\caption{Disappearance of the pinned UMs in the FM phase. (a) $V_{ANE}$ imaging of the vertical branch of a 3~$\mu$m-wide Hall cross of $\mathrm{Fe}_{0.49}\mathrm{Rh}_{0.51}$/Pt in the AF phase at 25~$^\circ$C. The average between images at $H_{app} = \pm 2$~kOe shows pinned UMs as in Fig.~1 of the main text. (b) The same branch in the FM phase at 115~$^\circ$C. The average of $H_{app} = \pm 2$~kOe shows that the pinned UMs in the AF phase disappear in the FM phase; residual contrast near the edges may reflect imperfect image alignment. (b,c) The horizontal branch of the same cross, which we image at 25~$^\circ$C (b) and 115$^\circ$C (c) to avoid artifacts from aligning two images. The pinned UMs that we observe at 25$^\circ$C disappear in the FM phase at 120$^\circ$C.}
\label{5050_hightemp}
\end{figure*}

In this section we demonstrate that the micron-size regions of positive and negative $V_{ANE}$ we image in the AF phase in FeRh vanish above $T_C$ in the FM phase. In Fig.~\ref{5050_hightemp}(a) we image the vertical branch of a 3 $\mu$m-wide Hall cross of $\mathrm{Fe}_{0.49}\mathrm{Rh}_{0.51}$/Pt, similar to the samples imaged in the main text. We first image at 25~$^\circ$C at $H_{app} = \pm$2.0~kOe applied along $x$ and take the average to show the pinned UMs. We then image at 115~$^\circ$C -- in the FM phase -- applying $H_{app} = \pm$2~kOe, shown in Fig.~6(b). Plotting the average at 110~$^\circ$C on the same scale as the average at 25~$^\circ$C shows that the structure of the pinned UMs at 25~$^\circ$C disappears in the FM phase. Contrast near the edges may arise from imperfect alignment between the two images. To avoid this issue, we image the horizontal branch of the same cross at 25~$^\circ$C in Fig.~\ref{5050_hightemp}(b), which shows pinned UMs similar in structure to those shown in Fig.~1 of the main text. We then image the horizontal branch at 120~$^\circ$C in \ref{5050_hightemp}(c), in the FM phase, with $H_{app}$ = 2 kOe along $x$. The contrast from UMs disappears because the FM moments are saturated along $x$ and we measure $m_y$ in the horizontal branch. 

\section{Magnetometry of FeRh/Pt}

\begin{figure}[!htb]
\centering
\includegraphics[scale=0.6]{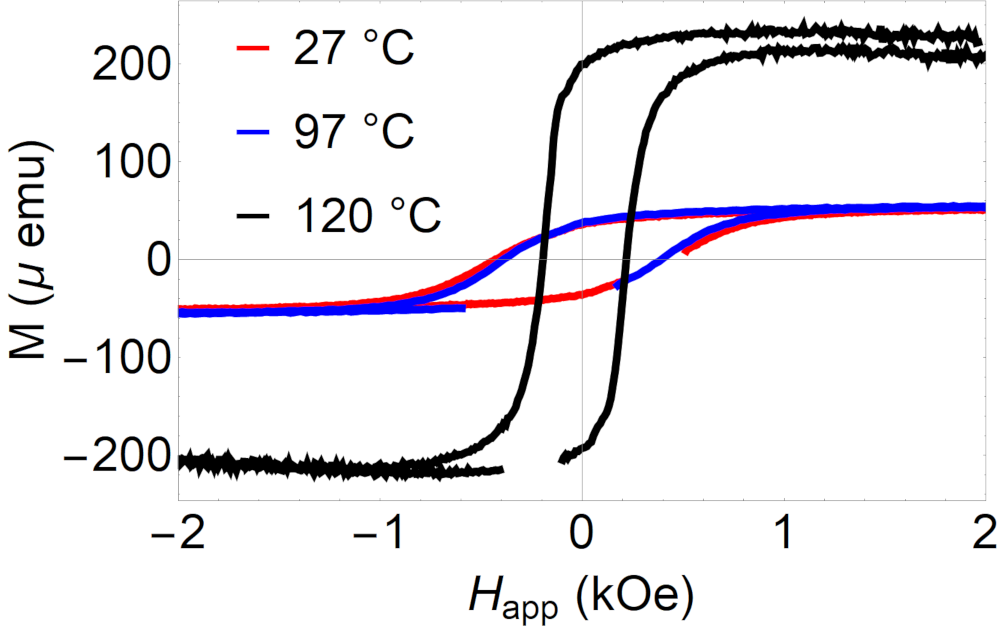}
\caption{VSM on Pt/FeRh. A weak residual moment with 1 kOe coercivity at 25~$^\circ$C is consistent with the unpinned UMs we measure with ANE imaging.}
\label{vsm_plot}
\end{figure}

We perform VSM measurements on an unpatterned film of $\mathrm{Fe}_{0.49}\mathrm{Rh}_{0.51}$/Pt using a Quantum Design Physical Property Measurement System (PPMS). In Fig.~\ref{vsm_plot} we plot $M(H)$ at 27~$^\circ$C, 97~$^\circ$C, and 120$~^\circ$C.

We observe characteristic FM hysteresis at 27~$^\circ$C with 750 Oe coercivity, which is consistent with the coercivity of the unpinned moments in the $V_{ANE}$ images. Although $V_{ANE}$ increases by a factor of 2 between 25~$^\circ$C and 70~$^\circ$C, the saturation moments at 25~$^\circ$C and 70~$^\circ$C are almost identical. The apparent discrepancy may be due to an increase in the ANE coefficient with increasing temperature \cite{NagaosaRevModPhys}.

\section{ANE imaging of varying FeRh stoichiometry}

In the main text we image 20 nm-thick $\mathrm{Fe}_{0.49}\mathrm{Rh}_{0.51}$ capped with Pt. In addition to ANE within the FeRh bulk, a longitudinal spin Seebeck effect (LSSE) at the Pt/FeRh interface could contribute to the $V_{ANE}$ voltage we measure, which would have the same symmetry as the ANE. To separate out any potential interfacial LSSE and explore the AF domain structure at different FeRh compositions, we image uncapped $\mathrm{Fe}_{0.43}\mathrm{Rh}_{0.57}$ and $\mathrm{Fe}_{0.52}\mathrm{Rh}_{0.48}$ in Fig.~8(a) and 8(b), respectively. Both samples are approximately 35 nm thick, grown on MgO(001) substrates. Note that while the FeRh/Pt films in the main text are sputtered, the $\mathrm{Fe}_{0.43}\mathrm{Rh}_{0.57}$ and $\mathrm{Fe}_{0.52}\mathrm{Rh}_{0.48}$ samples are grown by molecular-beam epitaxy. We image at positive and negative $H_{app}$ applied along $x$ and take the half-difference, following the same procedure as we do in the main text. We show both the horizontal and vertical branch of the $\mathrm{Fe}_{0.43}\mathrm{Rh}_{0.57}$ sample in the same image. 

\begin{figure*}[htb]
\centering
\includegraphics[scale=0.26]{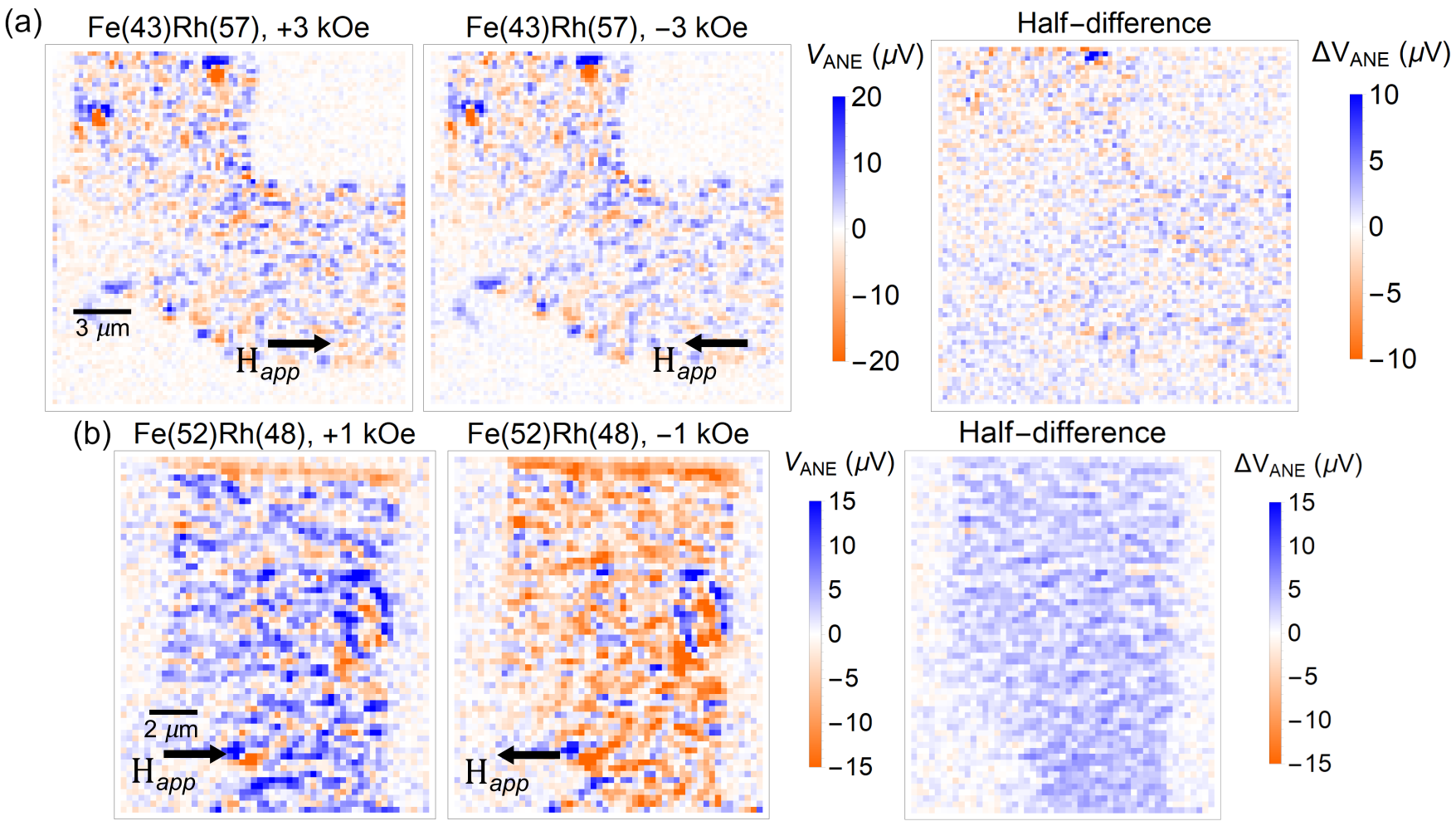}
\caption{ANE images of (a) $\mathrm{Fe}_{0.43}\mathrm{Rh}_{0.57}$ and (b) $\mathrm{Fe}_{0.52}\mathrm{Rh}_{0.48}$ at 25~$^\circ$C. Taking half-differences between positive and negative field shows unpinned moments in $\mathrm{Fe}_{0.52}\mathrm{Rh}_{0.48}$ and not in $\mathrm{Fe}_{0.43}\mathrm{Rh}_{0.57}$. This suggests that the unpinned UMs in $\mathrm{Fe}_{0.52}\mathrm{Rh}_{0.48}$ and $\mathrm{Fe}_{0.50}\mathrm{Rh}_{0.40}$/Pt are not due to a residual FM phase near the bottom FeRh/MgO interface, but arise instead from uncompensated excess Fe moments in the bulk. }
\label{stoichiometry}
\end{figure*}

\begin{figure*}[htb]
\centering
\includegraphics[scale=0.26]{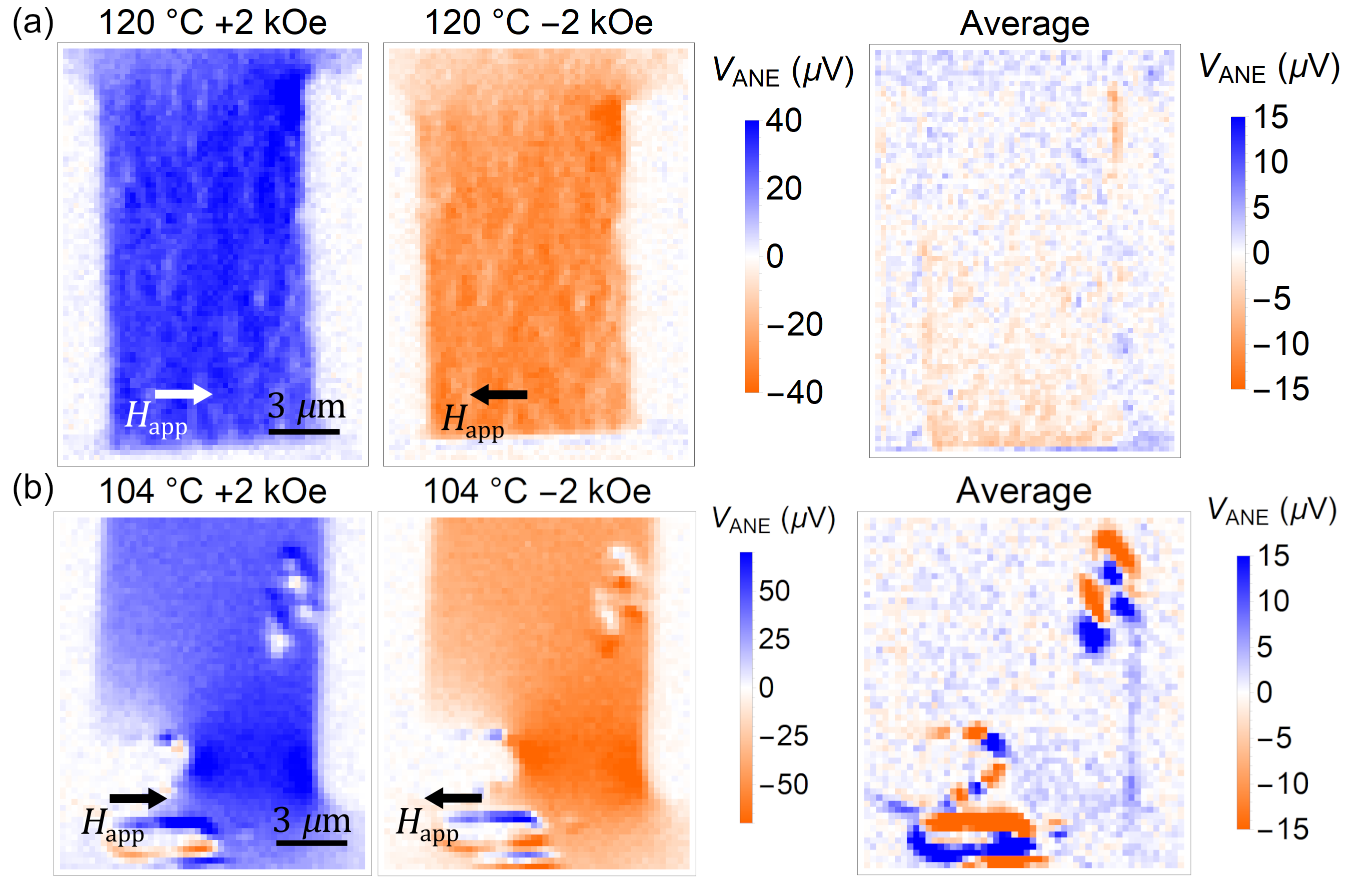}
\caption{Imaging the same (a) $\mathrm{Fe}_{0.43}\mathrm{Rh}_{0.57}$ and (b) $\mathrm{Fe}_{0.52}\mathrm{Rh}_{0.48}$ devices in the FM phase with $\pm H_{app}$ along $x$. Similar to the $\mathrm{Fe}_{0.50}\mathrm{Fe}_{0.50}$/Pt sample, the lack of contrast in the average image in the FM phase shows that the pinned UMs disappear above $T_C$ (although some dipolar artifacts are visible in (b) due to dirt on the sample surface).}
\label{high_temp_imaging}
\end{figure*}

Both $\mathrm{Fe}_{0.43}\mathrm{Rh}_{0.57}$ and $\mathrm{Fe}_{0.52}\mathrm{Rh}_{0.48}$ show submicron regions of positive and negative contrast similar in size, shape, and signal magnitude to the pinned UMs in the $\mathrm{Fe}_{0.50}\mathrm{Rh}_{0.50}$ samples. This indicates that any contribution from LSSE to the $V_{ANE}$ images of Pt/FeRh is smaller than the contribution from ANE. Interestingly, we observe both pinned and unpinned moments in the $\mathrm{Fe}_{0.52}\mathrm{Rh}_{0.48}$ sample, shown in the half-difference image, whereas we observe only pinned moments in the $\mathrm{Fe}_{0.43}\mathrm{Rh}_{0.57}$ sample. The unpinned UMs from uncapped $\mathrm{Fe}_{0.52}\mathrm{Rh}_{0.48}$ generate $V_{ANE}$ of similar magnitude as the unpinned UMs in $\mathrm{Fe}_{0.50}\mathrm{Rh}_{0.50}$/Pt, which indicates that the unpinned UMs in $\mathrm{Fe}_{0.50}\mathrm{Rh}_{0.50}$/Pt are most likely not caused by the Pt capping layer. In addition, our observation of unpinned UMs in $\mathrm{Fe}_{0.52}\mathrm{Rh}_{0.48}$ and not $\mathrm{Fe}_{0.43}\mathrm{Rh}_{0.57}$ is inconsistent with both strain and chemical diffusion-induced residual FM near the bottom FeRh/MgO interface, because both of these mechanisms predict more residual FM at higher Rh concentrations \cite{FanPRB}. Instead, our results are more consistent with unpinned UMs from excess Fe. Because perfect AF ordering of FeRh assumes exactly 50/50 stoichiometry, we expect the excess Fe atoms in the bulk at higher Fe concentration to be uncompensated. 

We check that $V_{ANE}$ in Fig.~8 is due to the N\'eel order and not spatial inhomogeneity in electrical resistance or sample quality by imaging the same devices in the FM phase in Fig.~9. In both $\mathrm{Fe}_{0.43}\mathrm{Rh}_{0.57}$ and $\mathrm{Fe}_{0.52}\mathrm{Rh}_{0.48}$, we observe nearly uniform FM and the inhomogeneous contrast in the AF phase disappears, seen by taking the average between images at positive and negative $H_{app}$ = 2 kOe. (Particles of dirt on the $\mathrm{Fe}_{0.52}\mathrm{Rh}_{0.48}$ sample produce non-magnetic dipole-like artifacts from the in-plane charge Seebeck effect).

\section{Effects of field-cooling on AF domains}

\begin{figure}[htb]
\centering
\includegraphics[scale=0.56]{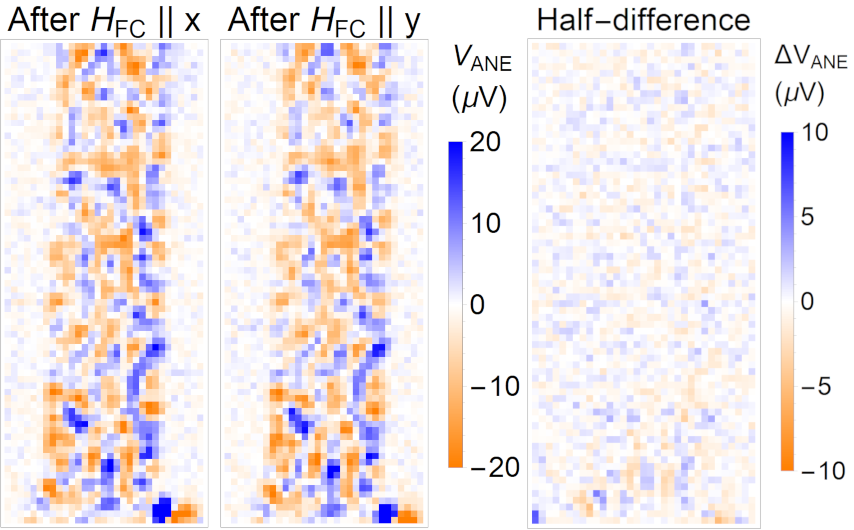}
\caption{Imaging the FeRh/Pt sample from Fig.~1 and Fig.~2 of the main text after cooling at $H_{FC}$ = 2 kOe, first along $x$ and then along $y$. The lack of contrast in the half-difference image indicates that we observe no effect of field-cooling on the structure of the AF UMs.}
\label{field_cooling}
\end{figure}

\begin{figure}[htb]
\centering
\includegraphics[scale=0.46]{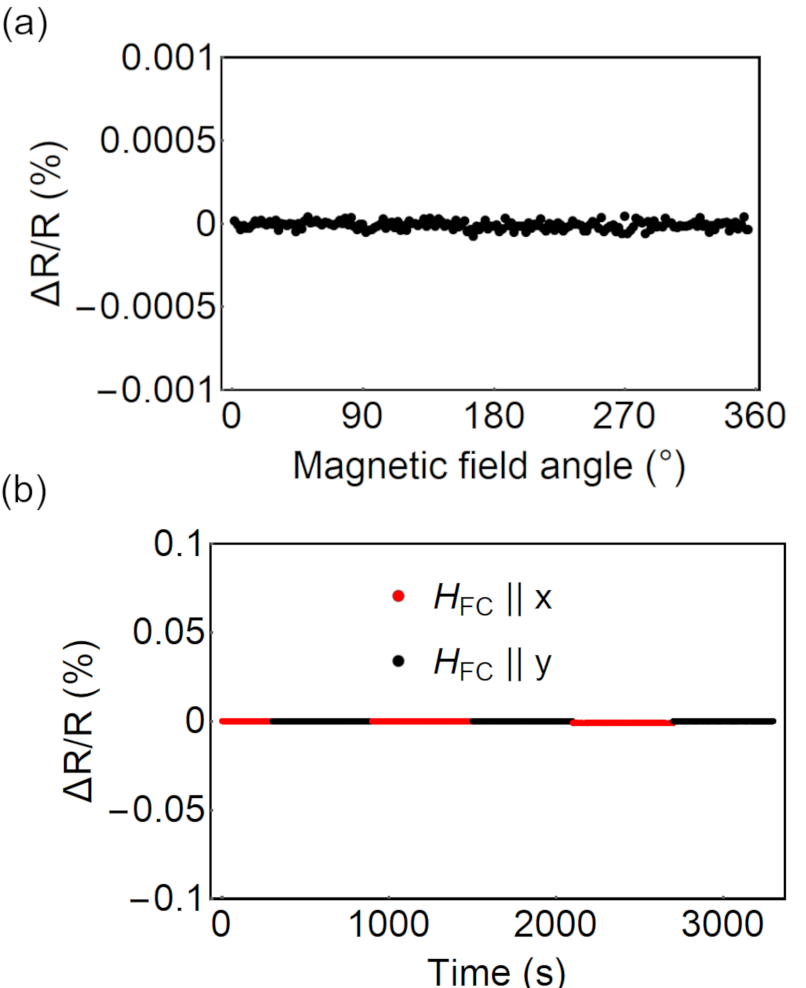}
\caption{Measuring the effects of field-cooling on the overall N\'eel orientation of another FeRh/Pt device using antiferromagnetic AMR. (a) Longitudinal resistance $R$ as a function of in-plane magnetic field angle at $H_{app}$ = 2 kOe. (b) Repeated measurements of $R$ while varying the direction of $H_{FC}$ by 90$^\circ$, which maximizes $\Delta R$ from AMR. We observe no $\Delta R$ in both cases, which means that both the pinned UMs and the AF domains are unaffected by field-cooling, at least at $H_{FC}$ = 2 kOe. }
\label{AMR}
\end{figure}

Previous reports on 50 nm-thick $\mathrm{Fe}_{0.50}\mathrm{Rh}_{0.50}$ \cite{MartiNatMater, MoriyamaAPL} showed reorientation of the bulk N\'eel order by field-cooling, which was measured using the antiferromagnetic anisotropic magnetoresistance (AMR). Pinned UMs in the AF phase are exchange-coupled to the N\'eel order, therefore if field-cooling reorients the AF domains in our samples we expect to observe changes in the room-temperature $V_{ANE}$ images. In Fig.~\ref{field_cooling} we first field-cool the device from Fig.~1 and 2 of the main text with $H_{FC}$ = 2 kOe along $x$, acquire an ANE image, then field-cool with $H_{FC}$ = 2 kOe along $y$ and acquire another ANE image.  Within our resolution and noise level, we observe no changes in the $V_{ANE}$ images after field-cooling, which is shown by the lack of contrast in the half-difference image. This is consistent with our findings in Appendix F that the pinned UM structure near $T_C$ is unaffected by $H_{FC}$. 
 
In Fig.~11, we measure antiferromagnetic AMR of another $\mathrm{Fe}_{0.50}\mathrm{Rh}_{0.50}$/Pt device from the same chip after field-cooling. We heat to the FM phase by Joule heating from DC current, following the procedure of Ref.~  \cite{MoriyamaAPL}, and measure the longitudinal resistance $R$ using a lock-in amplifier and a Wheatstone bridge. $R$ depends on the average N\'eel orientation $\vec{N}$ as $R = R_0 + \Delta R_{AMR}\cos^2 \theta$, where $\theta$ is the angle between $\vec{N}$ and the current density $\vec{j}$. We first measure $R$ at 25~$^\circ$C and $H_{app}$~=~2~kOe as a function of in-plane field angle in Fig.~\ref{AMR}(a) to measure AMR from any residual FM or uncompensated moments. In Fig.~\ref{AMR}(b) we then measure $R$ after repeatedly alternating $H_{FC}$ along $x$ and $y$, which maximizes $\Delta R$. We observe no $\Delta R$ in either case. From our noise level we place an upper bound of $\Delta R/R = 10^{-6}$ on any FM AMR at 25~$^\circ$C and $\Delta R/R = 10^{-5}$ on the maximum AF AMR. For comparison, the two existing studies on AF AMR in FeRh report $\Delta R/R = 1.7 \times 10^{-3} $ and $1.0 \times 10^{-4}$, respectively \cite{MartiNatMater, MoriyamaAPL}. Field-cooling therefore has no effect on the AF domain structure in our 20 nm-thick samples, which may be due to an increased effect of strain in our 20 nm-thick samples than in the 50 nm-thick samples used in the other studies.

\section{Antiparallel exchange coupling within the phase transition}

\textbf{\begin{figure*}[htb]
\centering
\includegraphics[scale=0.55]{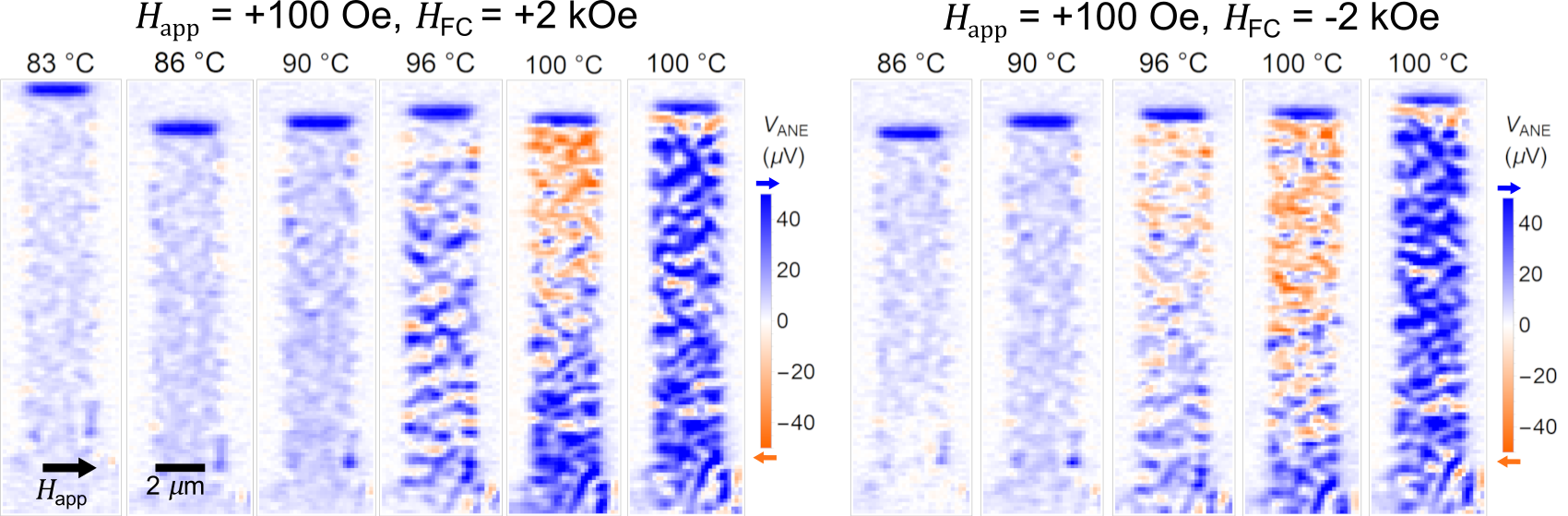}
\caption{Dependence of antiparallel exchange bias on the field-cooling field $H_{FC}$. (a,b) $V_{ANE}$ imaging after field-cooling  with $H_{FC}$ = +2 kOe (a) and -2 kOe (b) along $x$ to set the orientation of the AF UMs. We observe antiparallel exchange bias in both configurations, which means that $H_{FC}$ does not affect the orientation of the pinned UMs.}
\end{figure*}}

\begin{figure*}[htb]
\centering
\includegraphics[scale=0.63]{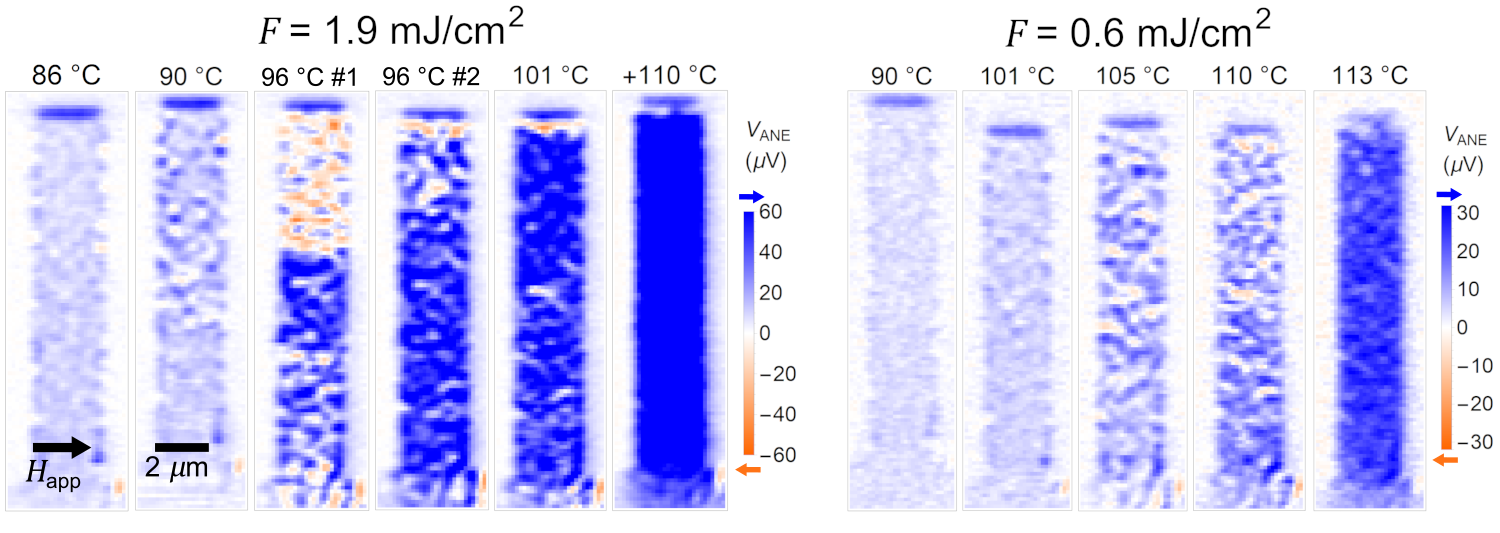}
\caption{Phase transition imaging at (a) laser fluence $F$ = 1.9~$\mathrm{mJ}/\mathrm{cm}^2$ and (b) $F$ = 0.6~$\mathrm{mJ}/\mathrm{cm}^2$.  We observe antiparallel exchange bias only at the higher fluence, which indicates that the repeated heating and cooling from the laser pulse train are necessary to induce a metastable frustrated spin state.}
\end{figure*}

In this section we investigate the dependence of exchange-biased emergent FM on the applied magnetic field $H_{app}$, the cooling field $H_{FC}$, and the laser fluence $F$. In conventional positive exchange bias in AF/FM multilayers, the orientation of the pinned AF UMs is set by field-cooling. Therefore, we first heat to the FM phase and then field-cool while applying $H_{FC}$ = +2 kOe along $x$ before imaging the increasing-temperature branch of the phase transition  in Fig.~12(a). We then field-cool at $H_{FC}$ = -2 kOe before imaging again in Fig.~12(b). In both sets of images we apply $H_{app}$ = +100 Oe along $x$, because it is larger than the 50 Oe coercivity field in the FM phase and smaller than the 1 kOe coercivity field of the unpinned UMs in the AF phase.

If the orientation of the pinned UMs is set by $H_{FC}$, we expect that they would be set along $+x$ in Fig.~12(a) and $-x$ in Fig.~12(b). Assuming an antiferromagnetic exchange coupling $J_{F-AF}$ between emergent FM and pinned UMs, we would therefore expect FM spins pinned antiparallel to $H_{app}$ only in Fig.~12(a) and not 12(b). Instead, we observe emergent FM spins antiparallel to $H_{app}$ in both cases. This result indicates that the pinned UMs are unaffected by field-cooling at least up to $H_{FC}$ = 2~kOe, which is further supported by the lack of any resolvable change in room-temperature $V_{ANE}$ images after field-cooling. 

After varying $H_{FC}$, we image the phase transition at laser fluence $F$ = $1.9~\mathrm{mJ}/\mathrm{cm}^2$ in Fig.~13(a), and then at $0.6~\mathrm{mJ}/\mathrm{cm}^2$. Both images are taken with $H_{FC}$ = +2~kOe and $H_{app}$ = +2~kOe along $x$. From the finite-element simulations of laser heating in Appendix G, we estimate the peak temperature increase at these fluences to be 16~$^\circ$C and 5~$^\circ$C, respectively. We observe antiparallel exchange bias only while using $F$ = 1.9~$\mathrm{mJ}/\mathrm{cm}^2$, which indicates that the pulsed laser has a different effect on the magnetic structure than the resistive heater we employ to adiabatically heat the whole sample. Imaging at fixed 93~$^\circ$C temperature as a function of fluence in Fig.~14 shows antiparallel exchange bias only at intermediate fluence -- 1.9~$\mathrm{mJ}/\mathrm{cm}^2$ and 2.2~$\mathrm{mJ}/\mathrm{cm}^2$. Higher fluences locally heat the FeRh into the FM phase, which persists after cooling because of the hysteresis of the 1st-order phase transition. This means that in the state of antiparallel exchange bias the sample locally remains in a mixed phase of AF and FM.

\begin{figure}[htb]
\centering
\includegraphics[scale=0.27]{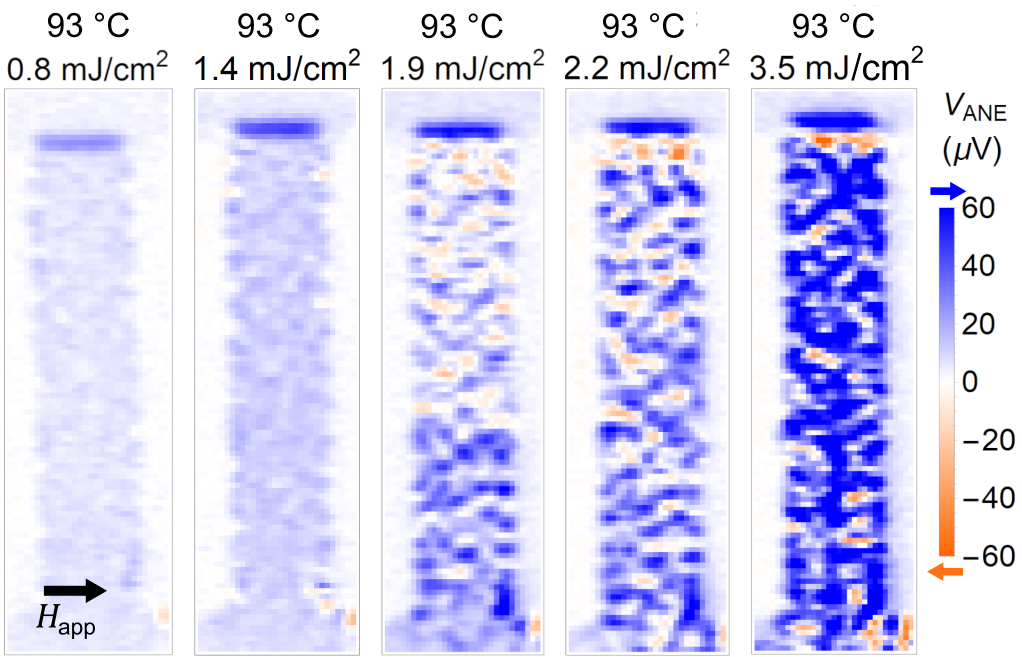}
\caption{Imaging at fixed $T$ = 93 $^\circ$C as a function of laser fluence $F$. We observe antiparallel exchange bias only at the intermediate fluences $F$ = 1.9~$\mathrm{mJ}/\mathrm{cm}^2$ and 2.2~$\mathrm{mJ}/\mathrm{cm}^2$. At lower fluences, the peak temperature increase is not high enough to unpin the pinned UMs. At higher fluence, the laser locally heats into the FM phase, which remains stable after cooling due to phase transition hysteresis.}
\end{figure}

Here we present one possible explanation. The laser locally heats the sample above the blocking temperature $T_B$ but below the transition temperature $T_C$, which unpins the pinned UMs without heating to the full FM phase. The Zeeman energy tends to rotate the newly unpinned UMs parallel to $H_{app}$ while $J_{F-AF}$ rotates the UMs antiparallel to $H_{app}$, resulting in a frustrated spin state. In addition, $J_{F-AF}$ weakens at higher temperature close to $T_C$, therefore the repeated heating and cooling from the laser pulses cause the newly unpinned UMs to undergo repeated reversals. Some of the UMs reorient parallel to $H_{app}$ before becoming repinned, in a similar process to training-induced positive exchange bias in AF/FM bilayers. After the UMs become repinned, they rotate the FM spins to be antiparallel to $H_{app}$. Note that we observe antiparallel exchange bias even at $H_{app}$~=~100~Oe, which means that the picture of a large $H_{app}$ overcoming $J_{F-AF}$ is not correct in this case. Instead, we must posit that when the pinned AF spins become unpinned, both the exchange coupling to the bulk AF order and the exchange coupling $J_{F-AF}$ to the emergent FM order are reduced, and when they become repinned both couplings increase again. 

\section{Finite-element simulations of laser heating}

\begin{figure*}[h]
\centering
\includegraphics[scale=0.46]{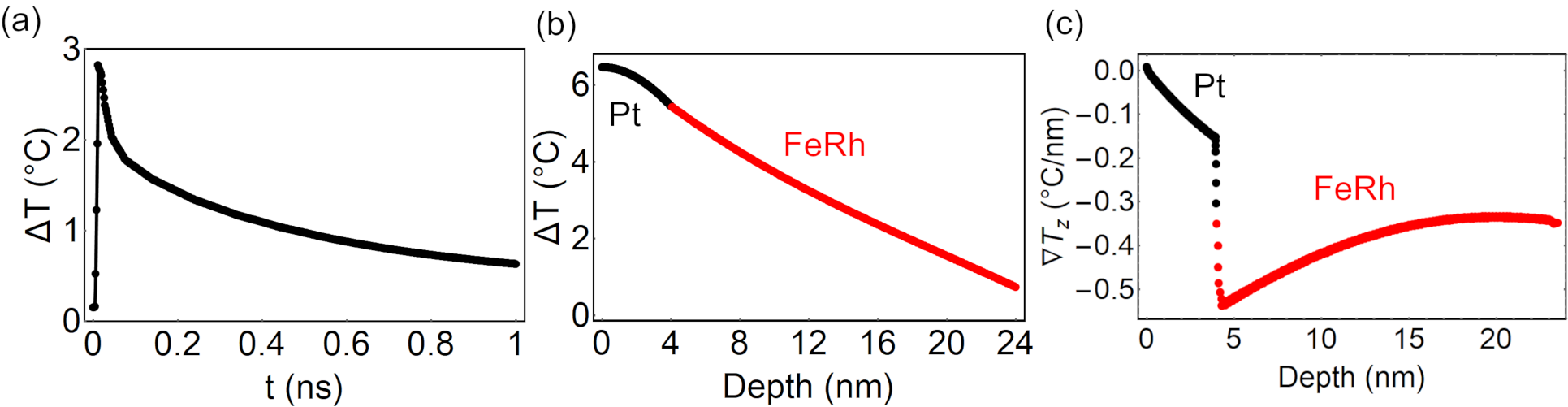}
\caption{Simulated laser heating profiles of 20 nm FeRh/4 nm Pt at $0.6~\mathrm{mJ}/\mathrm{cm}^2$ fluence. (a) Temperature increase $\Delta T$ as a function of time at a point in the center of the FeRh layer. (b) Depth profile of $\Delta T$ at peak heating at 20 ps, with Pt and FeRh layers indicated. (c) Thermal gradient profile $\nabla T_z (z)$ at 20 ps. }
\label{COMSOL}
\end{figure*}

We perform finite-element simulations of laser heating in 20 nm FeRh/4 nm Pt using the COMSOL Multiphysics\textsuperscript{\textregistered} software package. Representative simulation results at 0.6~$\mathrm{mJ}/\mathrm{cm}^2$ fluence are shown in Fig.~\ref{COMSOL}; assuming laser heating does not damage the sample, the temperature increases  linearly with fluence. We model the laser as a distributed heat source that exponentially decays according to the absorption depths of Pt and FeRh. Further details of the simulations are given in our previous work \cite{BartellNatComm}.

We plot temperature increase as a function of time $\Delta T(t)$ in the center of the FeRh layer in Fig.~\ref{COMSOL}(a), and temperature increase as a function of depth $\Delta T(z)$ at peak heating in Fig.~\ref{COMSOL}(b). We employ 3~$\mathrm{mJ}/\mathrm{cm}^2$ fluence in the ANE images in the main text and 0.6 and 1.9~$\mathrm{mJ}/\mathrm{cm}^2$ in the fluence-dependent images in Appendix F. These fluences cause peak temperature increases of 25~$^\circ$C, 5~$^\circ$C, and 16~$^\circ$C, respectively. 

We plot $\nabla T_z(z)$ at 0.6~$\mathrm{mJ}/\mathrm{cm}^2$ in Fig.~\ref{COMSOL}(c), which is just the derivative of $\Delta T(z)$ in \ref{COMSOL}(b). Although $T$ is continuous across the Pt/FeRh interface, $\nabla T_z$ is discontinuous due to the different thermal conductivity of Pt and FeRh. Note that $|\nabla T_z|$ is greater near the Pt/FeRh interface. Because $V_{ANE}$ is proportional to $m$ weighted by $\nabla T_z(z)$, we are slightly more sensitive to moments near the FeRh/Pt interface than near the MgO/FeRh interface.

\end{document}